\documentclass[prb,superscriptaddress,showpacs,twocolumn]{revtex4-1}
\usepackage{graphicx}
\usepackage{dcolumn}
\usepackage{bm}
\usepackage{amsmath}
\usepackage{amssymb}
\usepackage{ulem}
\usepackage{color}
\usepackage{float}

\newcommand{\ket}[1]{\left|#1\right\rangle}
\newcommand{\bra}[1]{\left\langle#1\right|}
\newcommand{\braket}[2]{\left\langle#1|#2\right\rangle}

\def\B{{\mathcal B}}
\def\E{{\mathcal E}}

\def\QM{{Q_{r_s}}}

\def\NL{{M_\Lambda^{}}}

\def\Z{\mathbb Z}

\def\V{{\mathcal V}}

\def\HF{{\rm HF}}

\def\HT{Hex$^{(2)}$ }

\def\Tr{{\bf Tr}}
\def\bQ{{\bf Q}}
\def\bq{{\bf q}}
\def\br{{\bf r}}
\def\bL{{\bf L}}
\def\bk{{\bf k}}
\def\bx{{\bf x}}

\def\be{{\bf e}}
\def\bT{{\bf T}}

\def\uo{\underline{1}}
\def\ut{\underline{2}}

\def\upuparrows{\uparrow\!\uparrow}

\def\downdownarrows{\downarrow\!\downarrow}

\begin{document}

\title{Properties of Hartree-Fock solutions of the three-dimensional electron gas}
\author{L. Baguet}
\affiliation{LPTMC, UMR 7600 of CNRS, Universit\'e P. et M. Curie, Paris, France}
\author{F. Delyon}
\affiliation{CPHT, UMR 7644 of CNRS, \'Ecole Polytechnique, Palaiseau, France}
\author{B. Bernu}
\affiliation{LPTMC, UMR 7600 of CNRS, Universit\'e P. et M. Curie, Paris, France}
\author{M. Holzmann}
\affiliation{LPTMC, UMR 7600 of CNRS, Universit\'e P. et M. Curie, Paris, France}
\affiliation{Univ. Grenoble Alpes, LPMMC, F-38000 Grenoble, France
\\
CNRS, LPMMC, F- 38000 Grenoble, France}

\date{\today}
\begin{abstract}
In a previous letter, L. Baguet et al., (Phys. Rev. Lett. {\bf 111}, 166402 (2013)), we presented the ground state phase diagram of the homogeneous electron gas in three dimensions within the Hartree-Fock approximation yielding incommensurate crystal states at high density. 
Here, we analyze the properties of these solutions. In particular, at high density we find universal behavior of the incommensurate crystal strongly supporting the existence of a spin density wave ground state.
\end{abstract}
\maketitle

\section{Introduction}

Most solid properties depend on the electron behavior, 
and one of the fundamental issues in solid state physics is the understanding of how electrons roam  in crystals.
Within the independent particle approximation, electronic properties are studied  considering
a model where electrons interact only with the positive ions of the crystal, but each electron remains independent of the others.  
Within this picture, insulators are characterized by completely filled bands, otherwise the state is metallic. The failure of band theory to predict insulating behavior and the 
role played by correlations due to  electron-electron interaction has been subject to intense research starting from the
early work of Mott \cite{Mott}.

In the limit where the positive charge of the ions in the crystal is smeared out uniformly through the whole system,
electronic correlations are separated from crystal field effects. In this so called jellium model of a solid, the electrons  interact with each others via the Coulomb interaction ($\sim 1/r$) and  global electroneutrality is insured by a uniform positive charge background.
Despite its simplicity, this model is directly relevant in a few cases, for example solid sodium \cite{Na}.

The Hartree-Fock (HF) approximation where electron-electron interactions are replaced by a simple self-consistent field  provides a first step to go beyond the independent electron approximation \cite{Ashcroft}. However, the resulting Hartree-Fock equations are non-local and non-linear. Even for the homogeneous electron gas model, the quantitative determination of the ground state phase diagram -- which depends only on density -- has been a challenging task\cite{Giulani,Needs,Shiwei,HF-2008,HF-2D}; large part of the HF phase diagram in three dimensions has been established only recently \cite{HF3DEG-letter}.  

Two possible phases of jellium have been widely investigated: the Fermi Gas (FG), where electrons are completely delocalized and form a uniform negative charge distribution, and the Wigner crystal (WC) where electrons are organized on a lattice. At high densities the kinetic energy dominates so the FG is believed to be the ground state, and for low densities, the long-range interaction makes electrons form the WC. 

However, 
already Wigner argued that the unpolarized FG is unstable even for high densities\cite{Wigner}. Later, Overhauser\cite{Overhauser} showed the instability of the unpolarized FG with respect to spin-density waves (SDW) within HF.

In Ref.\cite{HF3DEG-letter}, we have established the ground state phase diagram considering periodic states,
which, besides the WC, also include the possibility of incommensurate crystals (IC) where the number of maxima in the charge (or spin) density is larger than the number of electrons. At high density these calculations confirmed the FG instability with respect to IC.
In this paper, we provide a more complete description of both the method and the results. In particular, we focus on the physical properties of the HF states and on the universality of the results at high density giving further evidence for the SDW character of the high density ground state.

The paper is organized as follows. 
In Sec.\ref{SEC-MODEL}, we describe the model and introduce the basic definitions.
Some technical aspects on its numerical solutions are given in Sec.\ref{SEC-NUMDET}. 
The discussion of the results is presented in Sec.\ref{SEC-RES} with our conclusions in Sec.\ref{SEC-Conclusion}.

\section{The Model}\label{SEC-MODEL}
 
We consider a periodic system, with $N$ electrons of mass $m_e$ in a three dimensional box of volume $V$, embedded in an homogeneous background of opposite charge with a density $N/V=3/(4\pi a_B^3 r_s^3)$ where $a_B$ is the Bohr radius.
The Hamiltonian reads: 
\begin{align}
\label{Hamiltonian}
H=-\frac{\hbar^2}{2m_e} \sum_i \Delta_i+\frac1V \sum_{\bk\ne0}  v_\bk \sum_{i<j} e^{i\bk.\br_{ij}}
\end{align}
where $v_\bk=4\pi/\|\bk\|^2$.

Hartree-Fock solutions are exterior products (Slater determinants) of single particle states $\phi_\alpha$  denoted as $\ket{\Psi}=\bigwedge_{\alpha\in S} \ket{\phi_\alpha}$.
In terms of density matrix, the Hartree-Fock solutions can be defined by a 1-body density matrix  $\rho_1$  such that $\Tr \ \rho_1=1$ and $0\leq\rho_1\leq 1/N$, that is all eigenvalues are in $[0,1/N]$\cite{Coleman}. The two-body density matrix $\rho_2$ satisfies:
\begin{align}
\label{eq-rho2-rho1}
\rho_2(\uo,\ut;&\uo',\ut')=\rho_1(\uo;\uo')\rho_1(\ut;\ut')-\rho_1(\uo;\ut')\rho_1(\ut;\uo').
\end{align}
Now we  restrict our study to periodic states:  let $\Lambda^*$ be a lattice generated by $\bL_1,\bL_2$ and $\bL_3$, and $\rho_1(\br+\bL_i,\br'+\bL_i)=\rho_1(\br,\br')$.

The reciprocal lattice $\Lambda$ is generated by $\bQ_1,\bQ_2$ and $\bQ_3$ such as $\bL_i.\bQ_j=2\pi\delta_{ij}$ and we note $\B$ its Brillouin zone.
Then 
\begin{align}\label{BlockW}
\rho_1=\bigoplus_{\bk \in \B} \rho_\bk
\end{align}
where $\rho_\bk$ are positive matrices 
satisfying $0\leq\rho_\bk\leq 1/N$. 
The simulation box is a parallelepiped generated from the vectors $M\bL_i$, where $M$ is some integer. So $V\sim M^3$, and the number of $\rho_\bk$ is $M^3$. The thermodynamic limit is recovered for $M\rightarrow\infty$.
The total energy per electron in Hartree unit reads:
\begin{align}
\label{eq-etot-op}
	e&=\sum_{\bk\in\B} \Tr \left(K_\bk + \V_\bk\right) \rho_\bk\\
	K_\bk(\bq,\sigma;\bq',\sigma')&=\frac{a_B^2}{2}\|\bk+\bq\|^2\delta_{\bq\bq'}\delta_{\sigma\sigma'}
\end{align}
\begin{align}
\nonumber 
	&\V_{\bk_1}(\bq_1,\sigma_1;\bq_1',\sigma_1')=\frac {2\pi a_BN}{V} \times \\
\label{eq-Vk}
 		&\left[\delta_{\sigma_1\sigma_1'}  \sum_{\bq\in \Lambda} \frac{\delta_{\bq_1'-(\bq_1-\bq)}}{\|\bq\|^2}\right.
		\sum_{\substack{\bk_2 \in \B \\ \bq_2\in \Lambda \\ \sigma}}\rho_{\bk_2}(\bq_2,\sigma;\bq_2-\bq,\sigma)  \\
 \nonumber
&-\!\!\!\sum_{\substack{\bk_2 \in \B\\ \bq,\bq_2\in \Lambda}}\left.\frac{\delta_{\bq_1'-(\bq_1-\bq)}}{\|\bk_1+\bq_1-\bk_2-\bq_2\|^2} \rho_{\bk_2}(\bq_2,\sigma_1;\bq_2-\bq,\sigma_1')\right]
\end{align}
with $\Tr A_\bk\rho_\bk=\sum_{\bq\bq'\in\Lambda,\sigma\sigma'}A_\bk(\bq,\sigma;\bq',\sigma')\rho_\bk(\bq',\sigma';\bq,\sigma)$.

In this paper, $k_F$ denotes the Fermi wave vector depending on the gas polarization.
We have
\begin{align}
\label{eq-defkf}
	k_Fa_B=\frac\alpha{r_s}, \qquad \alpha=\left(\frac{9\pi}{2n_s}\right)^{1/3}
\end{align}
where $n_s=1$ for fully polarized gas (P) and $n_s=2$ for the  unpolarized gas (U). 
The one-body density matrices of the Fermi gas (FG) reads
\begin{align}
	\rho_\bk(\bq,\bq')&=\frac1N \delta_{\bq\bq'}\Theta(k_{F}-\|\bk+\bq\|)\qquad &(P)\\
	\rho_\bk(\bq\sigma,\bq'\sigma')&=\frac1N\delta_{\bq\bq'}\delta_{\sigma\sigma'}\Theta(k_{F}-\|\bk+\bq\|)  &(U)
\end{align}
while FG energies (per electron) are in Hartree units
\begin{align}
E_{FG}&=\frac{3k_F^2}{10}-\frac{3k_F}{4\pi}
\end{align}

On the other hand, in the Wigner crystal,  each $\rho_\bk$ is $1/N$ times a projector of rank $n_s$ (full band).

In order to describe solutions between FG and WC,
we search for a lattice $\Lambda$ and a density matrix $\rho_\bk$  such that the number of particle per unit cell is near $n_s$ (or some multiple of $n_s$ for non-Bravais lattices). 
For extremal states, the eigenvalues of $\rho_\bk$ must be exactly $0$ or $1/N$, 
the number of strictly positive eigenvalues
is not known a priori but is expected to fall between 0 and $2n_s$ (or some multiple of $2n_s$ for non-Bravais lattices).

In practice, the first $\NL$ $\bq$-vectors in $\Lambda$ are used (see 
Table \ref{TAB-LATTICE}).
Thus $\rho_\bk$ is a $n_s\NL\times n_s\NL$ matrix.
Using the representation
\begin{align}
\label{eq-rhokUDU}
	\rho_{\bk}=\sum_i D_{\bk,i}\ket{u_{\bk,i}}\bra{u_{\bk,i}}
\end{align}
where $\braket{u_{\bk,i}}{u_{\bk,j}}=\delta_{ij}$, 
the condition $0\leq\rho_\bk\leq1/N$ becomes $0\leq D_{\bk,i}\leq1/N$.

Notice that each $\ket{u_{\bk,i}}$ is a Bloch wave of band index $i$, and can be decomposed on the $n_sM_\Lambda$ states $\ket{\bk+\bq ;\sigma}$ where $\bq\in\Lambda$. So we note :
\begin{align}
	\ket{u_{\bk,i}}=\sum_{\substack{\bq\in\Lambda \\ \sigma}}a_{\bk,i}(\bq,\sigma)\ket{\bk+\bq \,; \sigma}
\label{eq-defu}
\end{align}
where $a_{\bk,i}(\bq,\sigma)$ are the unknown complex numbers ($\propto~\!\!\NL M^3$).
Imposing the polarization allows us to further reduce this number.

The next subsections give details on how the parameters are found using a descent method (\ref{SEC-DESCENT}), and how the energies are efficiently computed (\ref{SEC-FFT}).

\begin{figure}
\begin{center}
\includegraphics[scale=0.16]{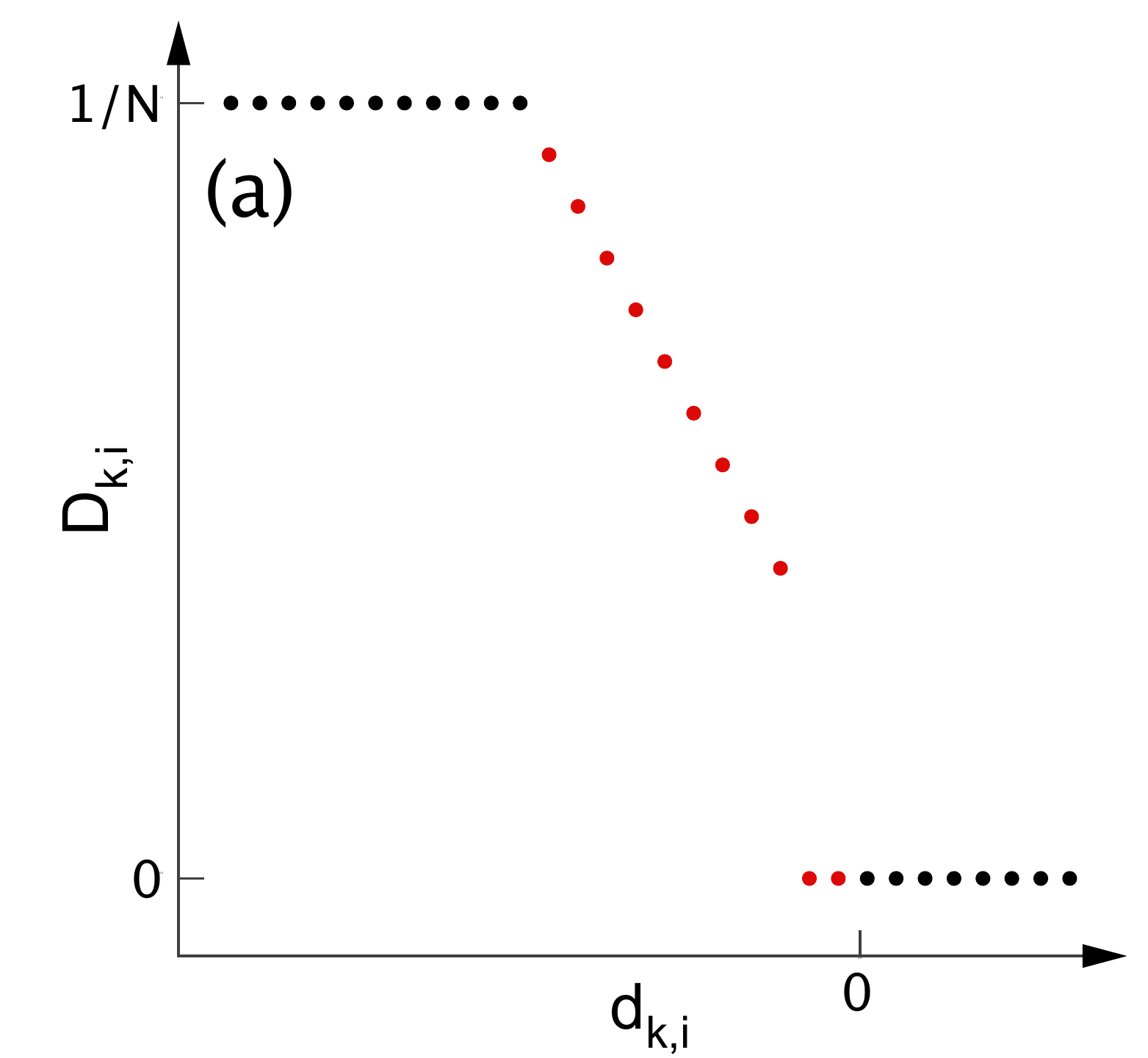}
\includegraphics[scale=0.16]{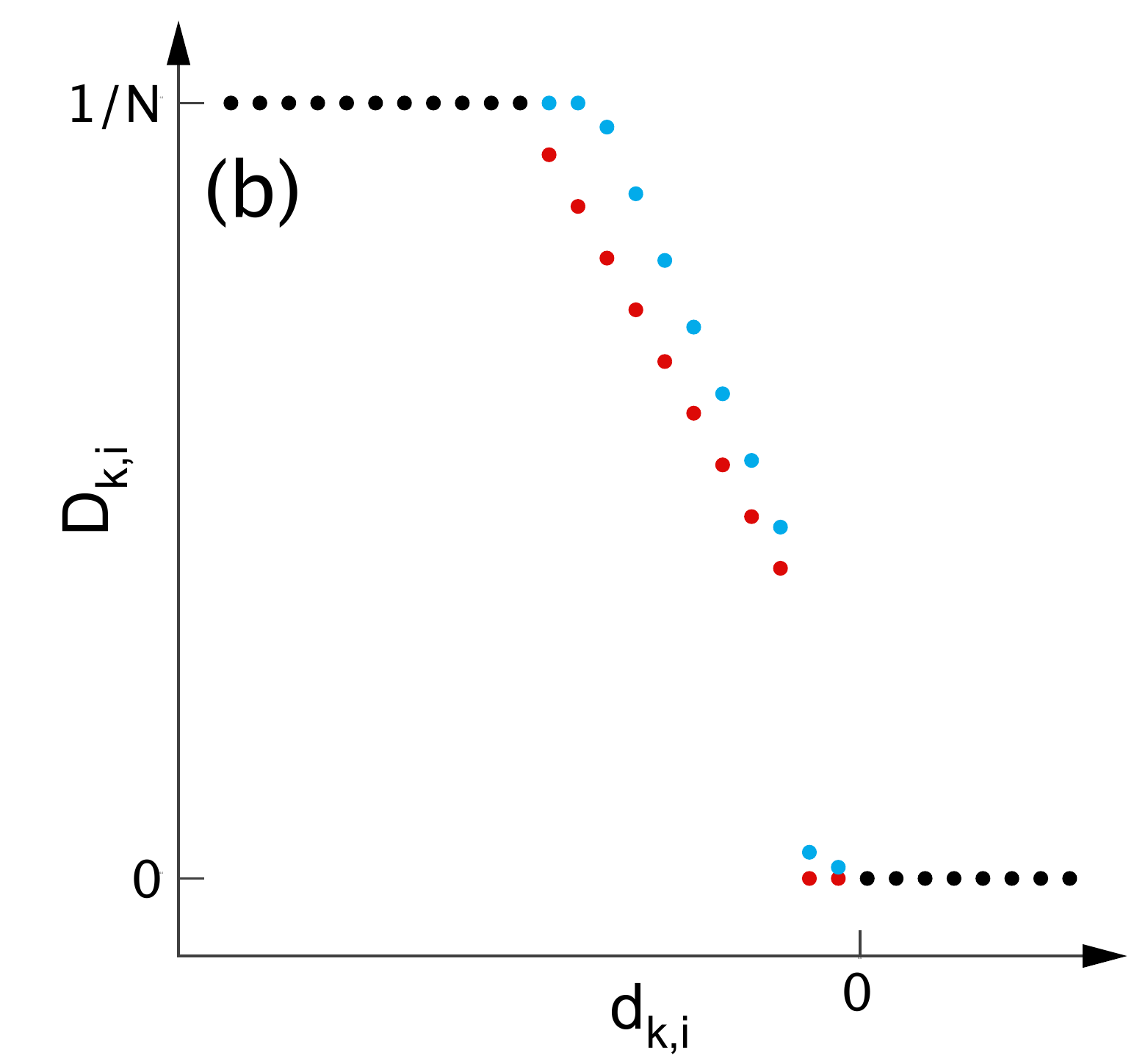}
\includegraphics[scale=0.16]{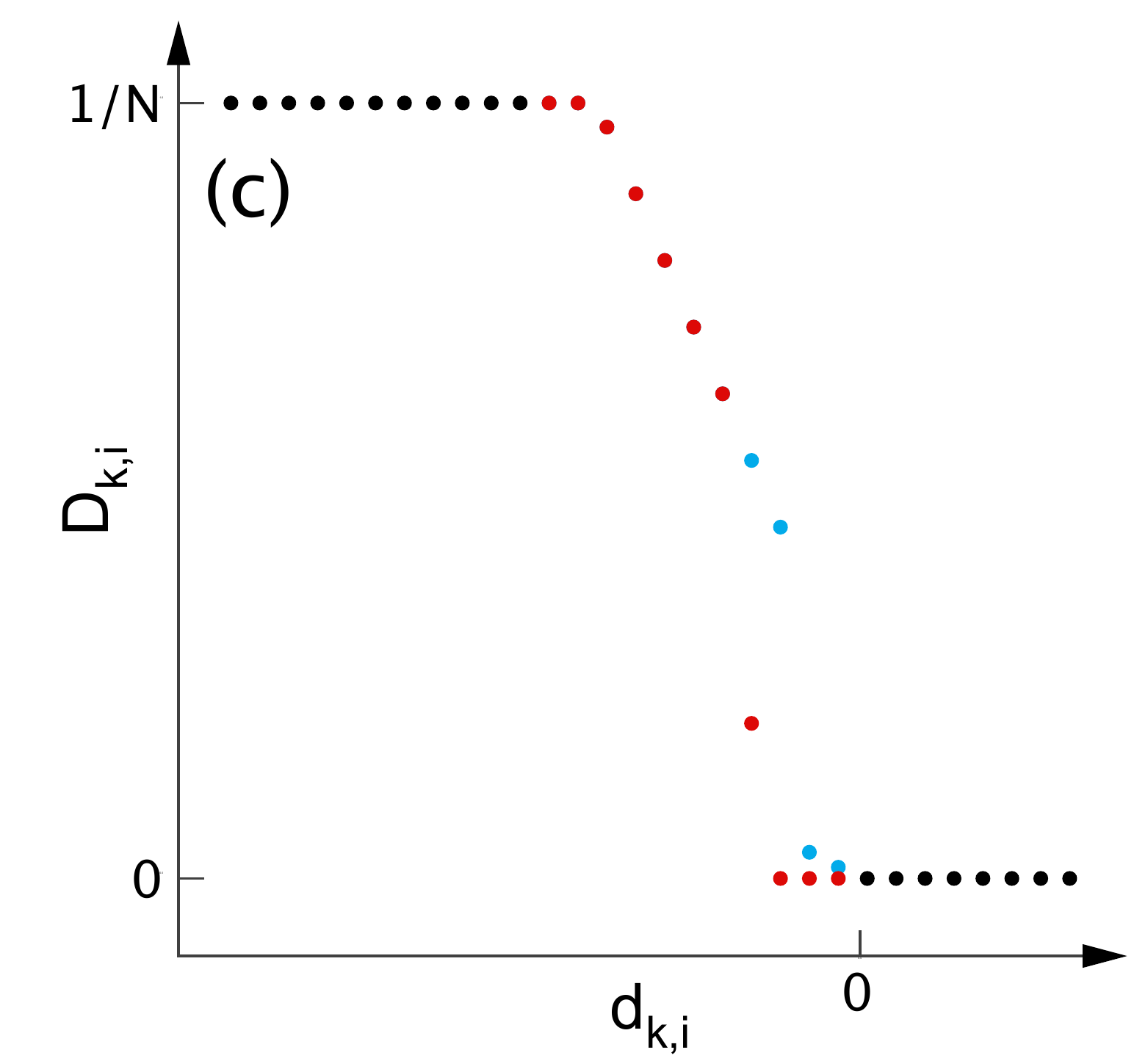}
\caption{
Illustration of the descent algorithm for the $D_{\bk,i}$ (see \ref{SEC-DESCENT}). 
(a): example of a set of $D_{\bk,i}$ and $d_{\bk,i}$ with $\sum_{\bk,i} D_{\bk,i}=1$.
Red points ($D_{\bk,i}<1/N$ and $d_{\bk,i}<0$) can move. 
(b): new set $D_{\bk,i}^{(\rm new)}$, see Eq.\ref{eq-Dnewset}, where
blue points have moved.
At this step, we have $\sum_{\bk,i} D_{\bk,i}^{(\rm new)}\ne1$ (here $>1$).
(c): corrected $D_{\bk,i}^{(\rm new)}$, in red, in order to insure that $\sum_{\bk,i} D_{\bk,i}^{(\rm new)}=1$.
\label{FIG-descentD}
}
\end{center}
\end{figure}

\begin{figure*}[t]
\begin{center}
\includegraphics[scale=0.5]{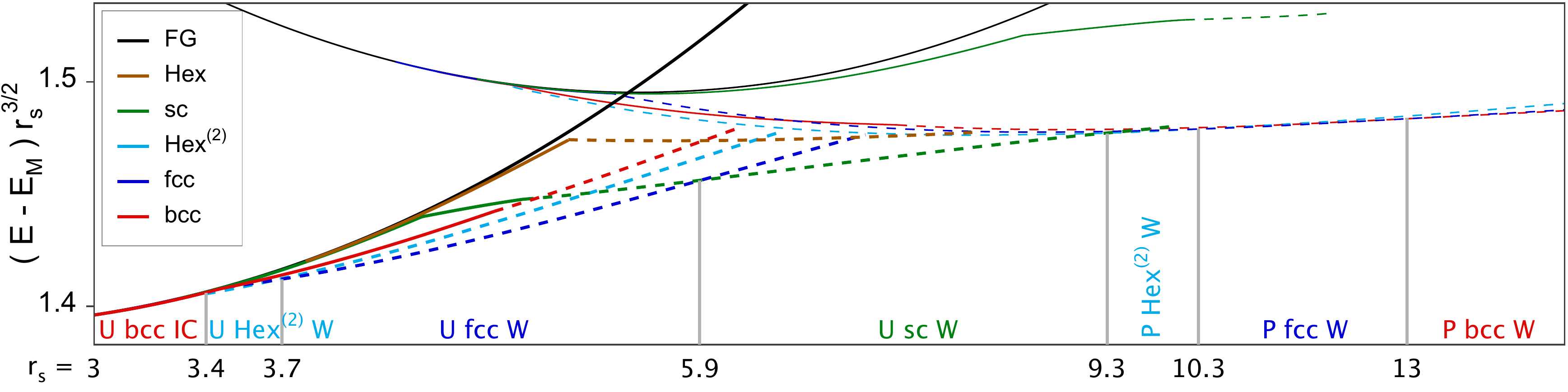}
\caption{
Hartree-Fock phase diagram of the 3D electron gas. 
Energies are in Hartree per electron. $E_M=-0.89593/r_s$ is the Madelung energy of a polarized-bcc Wigner crystal. Full lines stand for incommensurate regime ($Q>Q_W$) and dashed lines for the Wigner crystal ($Q=Q_W$). Thin lines stand for the polarized gas (upper curves) and thick lines for the unpolarized gas.\cite{LETT_SUPP}
\label{FIG-PhaDia3D}
}
\end{center}
\end{figure*}

\section{Numerical details}\label{SEC-NUMDET}

\subsection{Energy minimization}
\label{SEC-DESCENT}
From Eq.\,(\ref{eq-etot-op}) and Eq.\,(\ref{eq-rhokUDU}), the energy per electron and its variation read
\begin{align}
\label{eq-etot-u}
	e&=\sum_{\bk\in \B}\sum_i D_{\bk,i} \bra{u_{\bk,i}}\left( K_\bk+\V_\bk \right) \ket{u_{\bk,i}}\\
\label{eq-de}
	de&=\sum_{\bk\in \B}\Tr\, h_k^\HF d\rho_k
\end{align}
where 
\begin{align}
\label{eq-defHF}
	h_\bk^\HF&=K_\bk+2\V_\bk
\end{align}
is the so-called HF-Hamiltonian.\cite{Ashcroft,Giulani}
From
\begin{align}
\nonumber
	d\rho_k&=\ket{u_{\bk,i}}\bra{u_{\bk,i}} d D_{\bk,i} \\
		&\quad+  D_{\bk,i} (\ket{du_{\bk,i}}\bra{u_{\bk,i}}+\ket{u_{\bk,i}}\bra{du_{\bk,i}}),
\end{align}
Eq.(\ref{eq-de}) becomes
\begin{align}
	de=&\sum_{\bk\in \B}\left[d_{\bk,i} d D_{\bk,i}
		+2\Re \braket{ G_{\bk,i} }{ du_{\bk,i} }\right]
\end{align}
with
\begin{align}
\label{eq-grad-U}
G_{\bk,i}&=h_\bk^\HF\ket{u_{\bk,i}} \\
\label{eq-grad-D}
d_{\bk,i}&=\bra{u_{\bk,i}} h_\bk^\HF \ket{u_{\bk,i}}.
\end{align}

The minimization consists in the following steps: 
\begin{itemize}
\item[\it i)] choose $D_{\bk,i}$ and $\ket{u_{\bk,i}}$ to start with,
\item[\it ii)] for fixed $D_{\bk,i}$, find the best $\ket{u_{\bk,i}}$ with a quadratic descent method\cite{HF-2008},
\item[\it iii)] try to improve $D_{\bk,i}$ given $d_{\bk,i}$ and the linear constrains $0\le D_{\bk,i}\le1/N$ and  $\sum_{\bk,i} D_{\bk,i}=1$ and
in case of success, go to step {\it ii}. 
\end{itemize}
The process stops when each $D_{\bk,i}$ reaches its extrema 0 or $1/N$ with its gradient negative or positive, respectively.

Figure \ref{FIG-descentD} shows how the $D_{\bk,i}$ are moved. 
A new set is defined as 
\begin{align}
\label{eq-Dnewset}
	D_{\bk,i}^{(\rm new)}&=\max(0,\min(1/N,D_{\bk,i}-\varepsilon \,d_{\bk,i}))\\
	  t&=\sum_{\bk,i} D_{\bk,i}^{(\rm new)}    
\end{align}
If $t>1$,  some $D_{\bk,i}$ are decreased, those with the highest $d_{\bk,i}$ and $D_{\bk,i}>0$, as shown in Fig.\ref{FIG-descentD}-(c). 
Reversely, if $t<1$,  some $D_{\bk,i}$ are increased, those with the lowest $d_{\bk,i}$ and $D_{\bk,i}<1/N$.

Small $\varepsilon$ ($<0.1$) ensures that the $\ket{u_{\bk,i}}$ follow the $D_{\bk,i}$ {\sl adiabatically}. 
On the contrary, with large $\varepsilon$ ($\sim 1$), the system converge in a few steps, but the energy may end up in a local minimum.
An efficient compromise is to
start with $\varepsilon\sim1$, and decrease it at each step.
This allows a fast convergence to the same point as that obtained with  a small $\varepsilon$.
Except rare cases, all results are converged with 1 up to 30 moves of the $D_{\bk,i}$.

We checked on small system size ($M=8$ or 16) that the results do not depend on the starting point. 
Nevertheless, the speed of convergence can be significantly improved using conditioned initial states 
based on the description given in Sec. \ref{SEC-Characterization}.

\begin{figure}
\begin{center}
\includegraphics[scale=0.5]{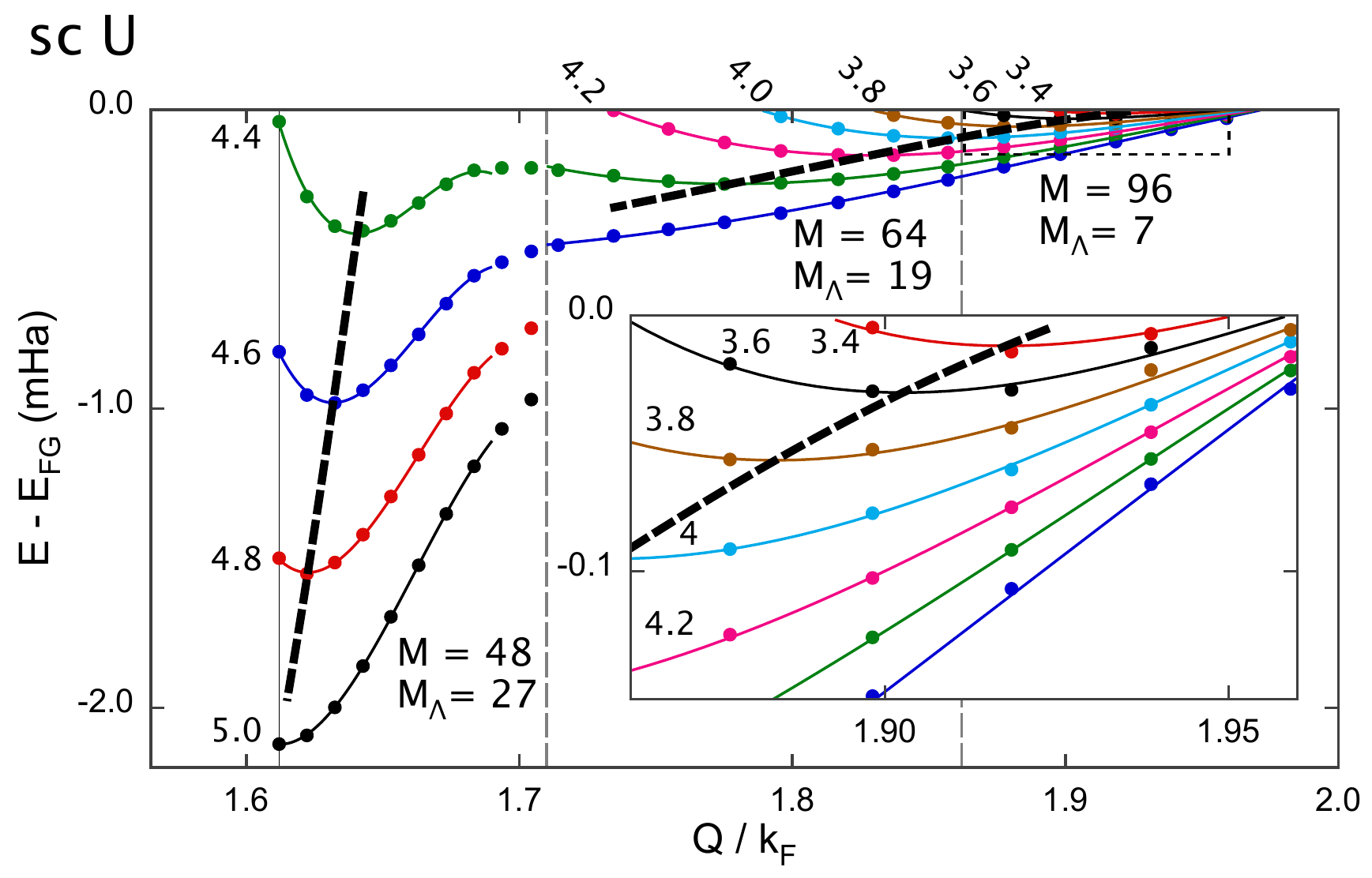}
\caption{
Energy versus the modulation $Q$ at various $r_s$ for the unpolarized gas in the sc symmetry.
Lines are the polynomial fits (see Eq. (\ref{eq-FIT})) of the numerical results (circles). 
$r_s$ is indicated at the start of each curve.
Thick dashed lines fit the minima of $\E(r_s,Q)$ at fixed $r_s$.
The leftmost vertical straight line stands for $Q=Q_W$.
Inset: zoom of the dotted rectangle of the main figure.
Gray dashed lines separate the domains of different $M$ and $M_\Lambda$ used in the numerics. The precision on the energy is always better than $10^{-5}$ Ha.
\label{FIG-ERSQ-CUBIC}
}
\end{center}
\end{figure}

\subsection{Implementation of $\V_\bk$ through FFT}
\label{SEC-FFT}

The largest computer time is the evaluation of $\V_\bk$ (Eq.\ref{eq-Vk}).
The computation of the exchange part of $\V_\bk$ involves a convolution over the Brillouin zone $\B$:
\begin{align}
h(\bk_1)=\sum_{\bk_2\in \B} \frac{1}{\|\bk_1-\bk_2\|^2} g(\bk_2)
\end{align}
Instead of computing $M^3$ terms ($\bk_1$), each of them containing a sum over $M^3$ terms ($\bk_2$), we would like to  implement the FFT over the Brillouin zone.
The main complication is that $\B$ is not a parallelepiped box (except in cubic case). 
Numerically we use the unit cell $B$ to index the functions $h$ and $g$.
\begin{align}
B = \left\{\bk=\sum_{\alpha=1}^3 \frac{n_\alpha}{M}\bQ_\alpha , \ 0\leq n_\alpha<M\right\} 
\end{align}
Any $\bk$ in $B$ corresponds to a unique $\hat \bk$ in  $\B$ with $\bk-\hat\bk \in \Lambda$.
With this notation we have to compute:
\begin{align}
h(\bk_1)=\sum_{\bk_2\in \B}f(\hat\bk_1-\hat\bk_2)g(\bk_2)
\end{align}
The function $f(\hat k_1-\hat k_2)$, may be seen as a function on:
\begin{align}
B_2&=\left\{\bk=\sum_{\alpha=1}^3 \frac{n_\alpha}M \bQ_\alpha, \ 0\leq n_\alpha<2M\right\}
\end{align} 
The dual $B_2^*$ is the $\Z/2M$ $\Z$-module generated by $\{\bL_\alpha/2\}$.
Introducing the ensemble of 8 vectors $E=\{\be=\sum_\alpha n_\alpha \bL_\alpha/2,\ n_\alpha=0 {\rm\ or\ }1\}$, we note $B^*_2=\cup_\be (B^*+\be)$ and $B^*$ is  $\Z/M$ $\Z$-module generated by $\{\bL_\alpha\}$. Thus we have:
\begin{align}
\label {conv2}
	f(\hat \bk_1-\hat \bk_2)&=\frac 1{ M^{3/2}}\sum_{\be\in E}\sum_{\bx\in B^*}\tilde f_\be(x)e^{-i(\hat \bk_1-\hat \bk_2)(\bx+\be)}\\
\label {conv3}
	\tilde f_\be(\bx)&=\frac 1{ 8M^{3/2}}\sum_{\bk\in {B}_2}f( \bk)e^{i\bk(\bx+\be)}
\end{align} 
Now we set $g_\be(\bk):=e^{i\hat \bk\be}g(\bk)$. Then using Eq.\ref{conv2}:
\begin{align}
\nonumber
f*g(\bk_1)&=\frac 1{M^{3/2}}\sum_{\be\in E}\sum_{\bx\in B^*}\tilde f_\be(\bx)e^{-i\bk_1\bx} e^{-i\hat \bk_1\be}\\
	&\qquad\quad\times \sum_{\bk_2\in B} g(\bk_2)e^{i\bk_2\bx} e^{i\hat \bk_2\be}\\
\label{prod} 
&=\sum_{\be\in E} e^{-i\hat \bk_1\be}\sum_{\bx\in B^*}\tilde f_\be(\bx)\tilde g_\be(\bx)e^{-i\bk_1\bx}
\end{align}
where 
\begin{align}
\tilde g_\be(\bx)=\frac 1{ M^{3/2}}\sum_{\bk\in \B}g_\be(\bk_2)e^{i\bk\bx}
\end{align}
 is the Fourier transform of $g_\be$.

Thus, once for all, the functions $\tilde{f}_\be(\bx)$ and $e^{i\hat \bk\be}$ are tabulated. Thereafter, at each step, $\tilde g_\be(\bx)$ are computed through eight FFTs.
Then the inverse FFT is applied to the eight products  $\tilde{f}_\be\tilde g_\be$. Finally, $h$ is obtained by summing the results with the weights $e^{-i\hat \bk_1\be}$.
Thus this  procedure is of order $M^3\ln M$ instead of $M^6$.

\subsection{Size effects}

In this sub-section we discuss the convergence of the results with respect to $\NL$ and $M$.

A finite $\NL$ is equivalent to a truncation of the Hilbert space (see Eq.\ref{eq-defu}), thus increasing $M_\Lambda$ leads to a lower energy.

At small $r_s$, the coefficients $a_\bk(\bq)$, in Eq.\ref{eq-defu}, decrease quickly with $\|\bk+\bq\|$.
On the contrary, as $r_s$ increases, the convergence is much slower.
Fortunately, it depends very weakly on $M$, allowing to extrapolate to large $\NL$ 
independently from $M$.

The extrapolation $M \to \infty$ corresponds to the thermodynamic limit. As shown in Ref.\cite{HF3DEG-letter},
the most important finite size effects can be written as
\begin{align}
\label{EQ-EM1}
	\Delta E_M \equiv E_M-E_\infty= \frac{E_1}{M} + \frac{E_2}{M^2} + \frac{E_3}{M^3}+...
\end{align}
where $E_1$ is related to the Madelung energy, and $E_2$ can be evaluated from the structure factor $S(\bk)$.
$E_3$ is non-analytical in IC but regular in WC allowing us a clean extrapolation 
to the thermodynamical limit.

The overall accuracy of our results depends both on $\NL$ and $M$. 
At large $r_s$, $\rho_\bk$ is smooth but extended, so $M_\Lambda$ must be as large as possible but not $M$. 
At small $r_s$, $\rho_\bk$ varies rapidly around the boundaries of the first $\B$, so $M$ must be as large as possible but not $\NL$.
Figure \ref{FIG-ERSQ-CUBIC} shows such an optimal compromise where the error on the total energy is always smaller than $10^{-5}$ Ha.

\section{Results}
\label{SEC-RES}

In Fig.\ref{FIG-PhaDia3D}. we recall the phase diagram\cite{HF3DEG-letter} displaying the geometry and polarization which yields the lowest energy at each value of $r_s$.
At fixed geometry, for large $r_s$, 
we find a Wigner commensurate crystal (WC), as expected. 
As $r_s$ decreases, an incommensurate metallic phase is found characterized by a modulation increasing from $Q_W$ to $2k_F$ as $r_s$ goes to 0.
Such an incommensurate crystal (IC) interpolates between the WC phase and the Fermi gas. 
Indeed, in the reciprocal space, the momentum distribution $n(\bk)$ evolves from that of the WC phase where it is a continuous function to the Fermi gas, where $n(\bk)=1$ for $\|\bk\|\leq k_F$ and 0 elsewhere. In IC states, increasing pockets are build around the corners of the $\B$ where $n(\bk)=0$.
Thus, in some directions $n(\bk)$ is continuous at the border of $\B$, as in the WC phase, whereas in other directions it is discontinuous, as in the FG at $k_F$.
In real space, an IC-state looks like a WC but with a larger number of lattice sites than the number of electrons.
At small $r_s$, for unpolarized states, the IC-states present a contrast of the charge density much smaller than that of the spin density.

In the next subsection we recall how the phase diagram is obtained and we establish the universality of the results at small $r_s$.
Subsection \ref{SEC-GAP} discusses the metal-insulator transition.
The subsection \ref{SEC-Characterization} is devoted to ground state wave function characterization.
Correlations are presented in the last subsection \ref{SEC-Gofr-Sk}.

\subsection{Ground state at fixed symmetry and polarization}
\label{SEC-GS-FIXEDSYM}

For each lattice symmetry, we compute the ground state $e(r_s,Q)$,  see for example Fig.\ref{FIG-ERSQ-CUBIC}.
Relevant values of $Q$ lie between $Q_W$ and $2k_F$, where $Q_W$ is the modulation of the Wigner crystal (see Table \ref{TAB-LATTICE}). 
At large $r_s$, the minimum of $e(r_s,Q)$ is at $Q_W$. 
As $r_s$ decreases, the lowest energy may be for $Q>Q_W$. 
In the neighborhood of a minima, we fit $e(r_s,Q)$ with a polynomial:
\begin{eqnarray}
\label{eq-FIT}
	e(r_s,Q)=\sum_{i=0}^{i_m}\sum_{j=0}^3 a_{ij} r_s^i Q^j,
\end{eqnarray}
where $i_m$ is generally 2 (and 1 when only few points are available).
For fixed $r_s$, $\QM$ minimizes $e(r_s,Q)$.
The thick dashed lines in Fig.\ref{FIG-ERSQ-CUBIC} is $e(r_s,\QM)$ as a function of $\QM$.
In Fig.\ref{FIG-PhaDia3D} are reported the dashed lines obtained for all geometries.

It is worth noticing that $\QM$ is always sufficiently large such that only the first band is occupied. 

\begin{figure}
\begin{center}
\includegraphics[scale=0.5]{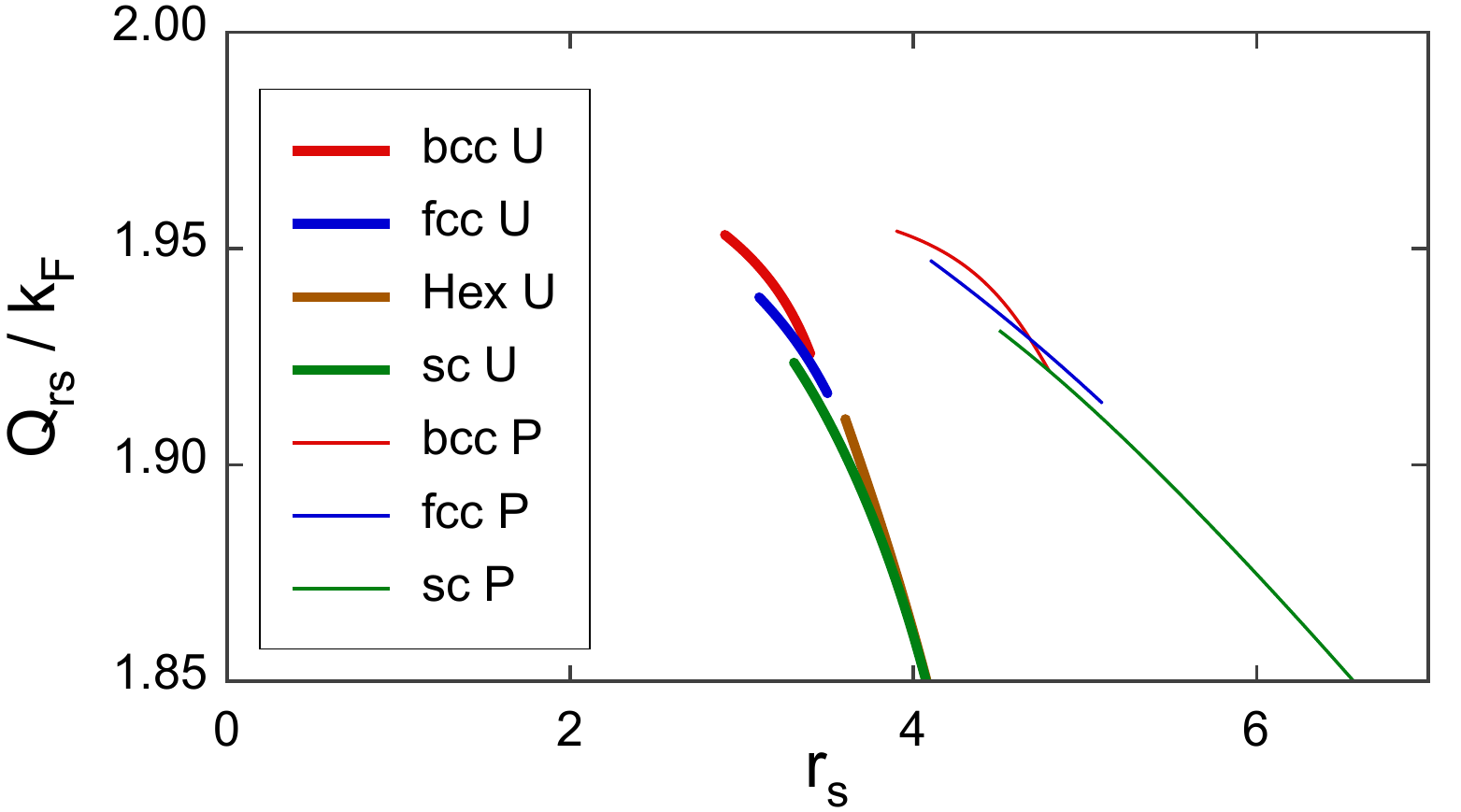}
\caption{
$\QM$ per symmetry and polarization, versus $r_s$. 
Data are obtained from fits (see text : Eq.\ref{eq-FIT}).
\label{FIG-QRS}
}
\end{center}
\end{figure}

Fig.\ref{FIG-QRS} shows $\QM$ versus $r_s$. 
As the FG is the ground state at $r_s=0$ and the Fermi surface is contained in $\B$
for $Q \ge 2 k_F$, we expect that these curves should reach $2 k_F$ for vanishing $r_s$,
compatible with our results.
Remarkably, at fixed polarization, $\QM$ slightly depends on the geometry.

\begin{figure}
\begin{center}
\includegraphics[scale=0.25]{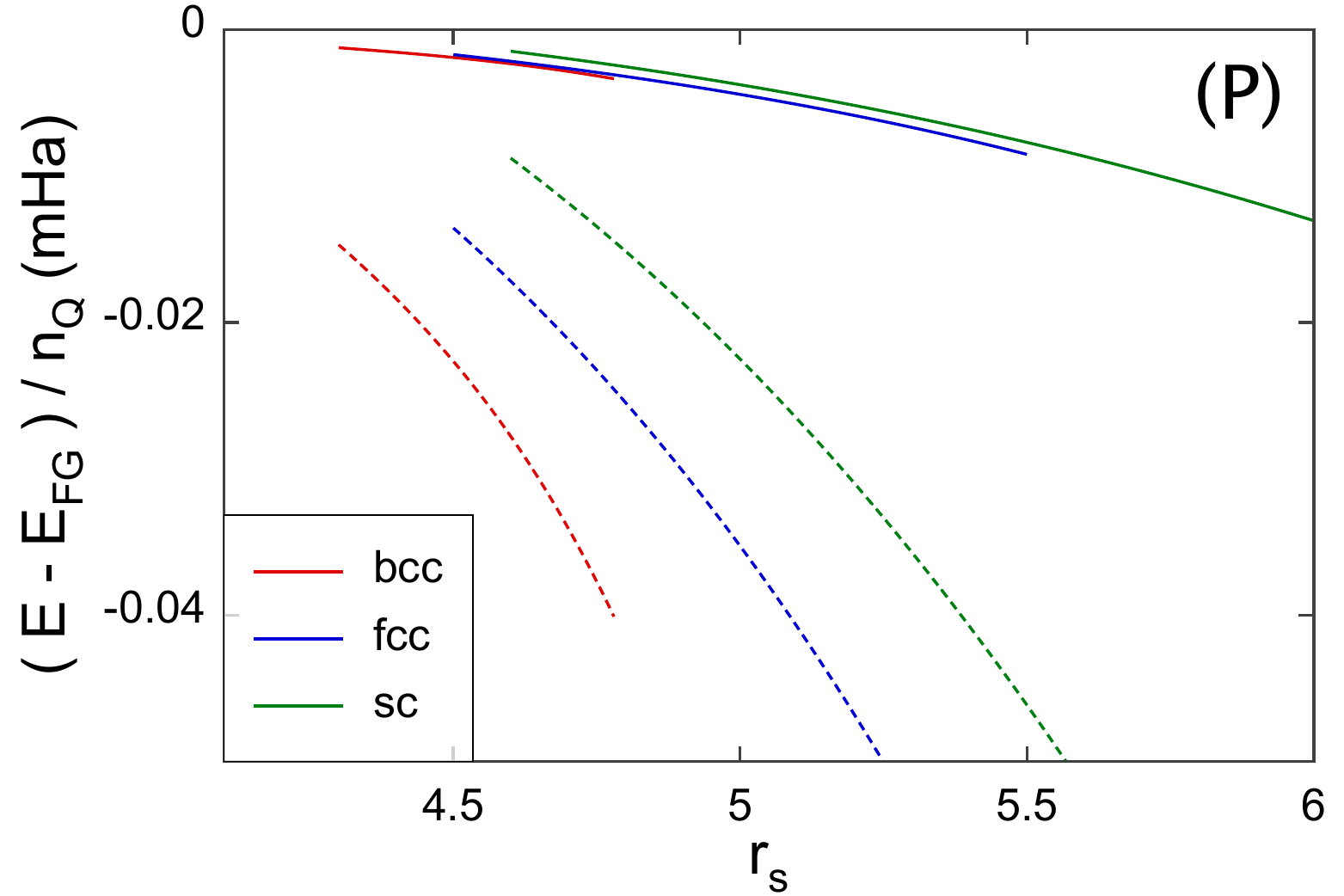}
\includegraphics[scale=0.25]{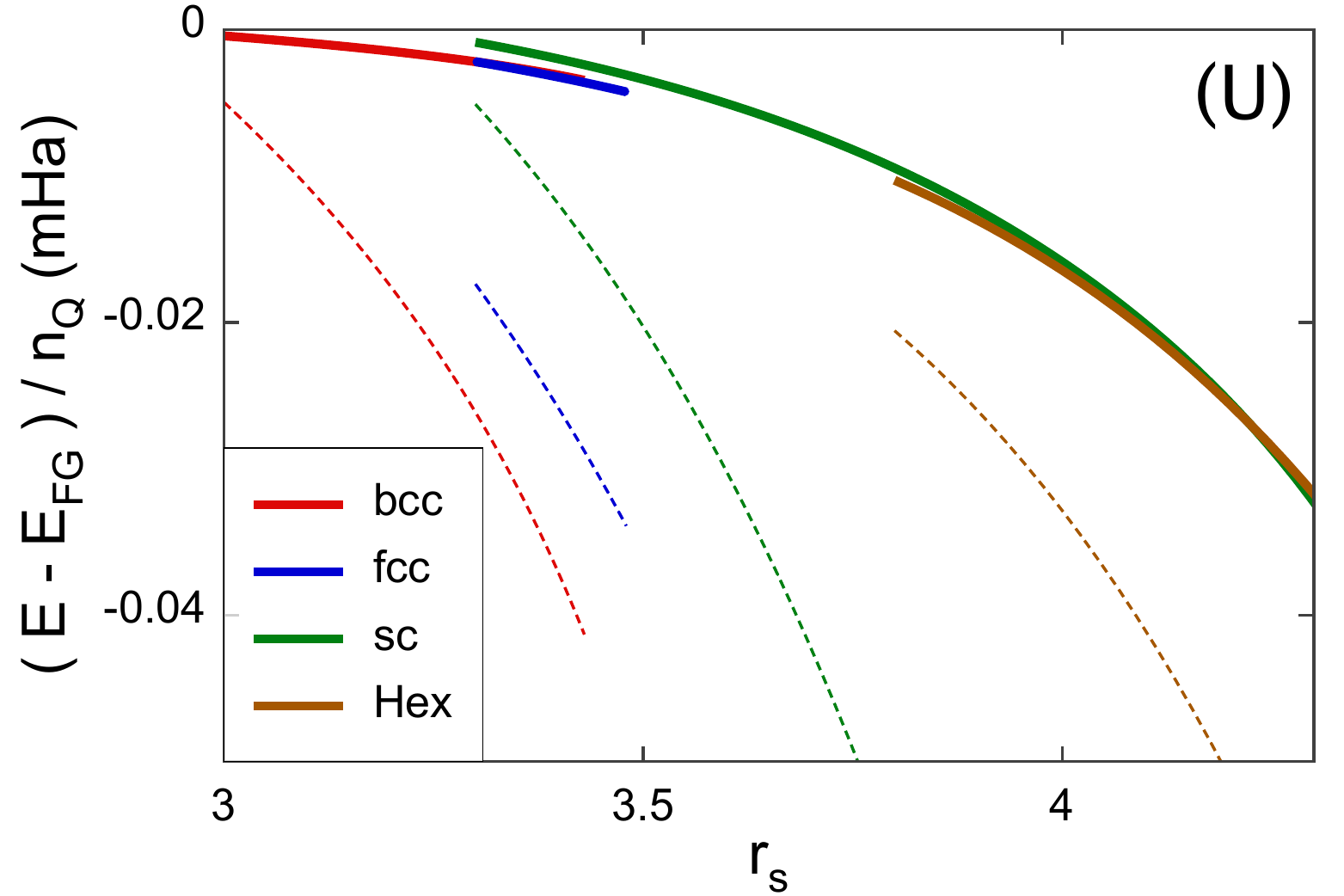}
\caption{
Energies versus $r_s$ of polarized (P) and unpolarized (U) IC. 
Dashed lines: raw energies.
Full lines: energies divided by the number $n_\bQ$ of available SDW's: $n_\bQ=2,6,8,12$ for Hex, sc, fcc, bcc, respectively.
\label{FIG-PhaDiaScaled}
}
\end{center}
\end{figure}

This can be understood assuming SDW/CDW\cite{Overhauser,Kurth} holds at small $r_s$.
A SDW is defined as a superposition of waves $\ket{\bk}$ and $\ket{\bk+\bQ}$, where $\bQ$ is a piecewise constant wave vector,  such that $\bk$ and $\bk+\bQ$ are close to the Fermi surface.
The energy gain of the SDW's is proportional to the number of available $\bQ$-vectors, assuming that the SDW's are independent.
If the $\Z$-module generated by the $\bQ$'s is discrete, then the model is actually periodic.
The number $n_\bQ$ of available $\bQ$ is the number of nearest neighbors of the origin in $\Lambda$, i.e. the number of points in the first shell of $\Lambda$.
Fig.\ref{FIG-PhaDiaScaled} shows 
that, once rescaled by $n_\bQ$, these energies 
become very close.
Whereas such a behavior may be expected at $r_s\ll1$, it is quite remarkable that it can be applied up to $r_s \approx 3$.
Further, it explains that bcc symmetry is favored as it has the largest  value $n_\bQ=12$. 

\subsection{Metal-Insulator transition}
\label{SEC-GAP}
\begin{figure*}
\begin{center}
\includegraphics[scale=0.42]{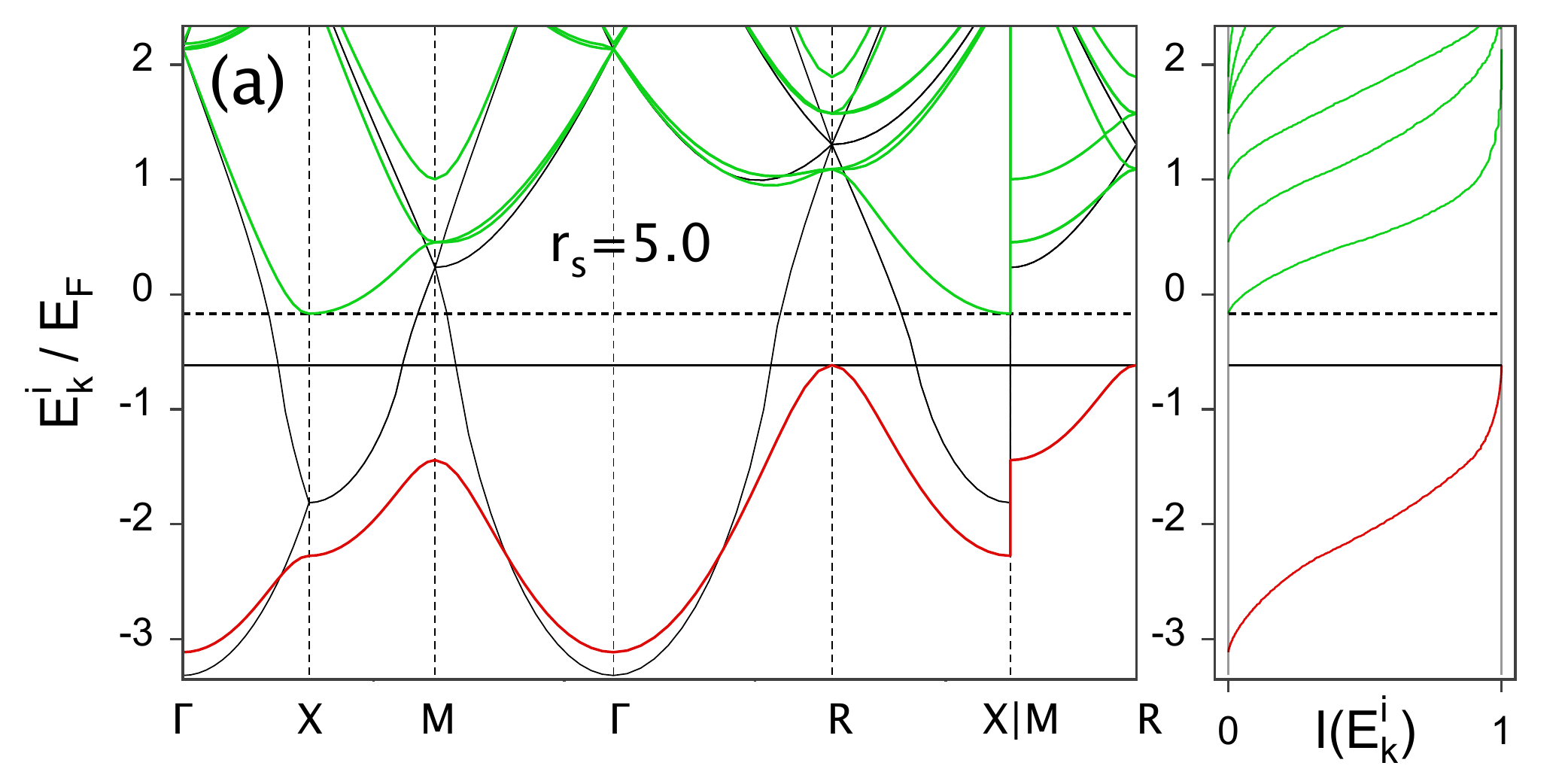}
\includegraphics[scale=0.42]{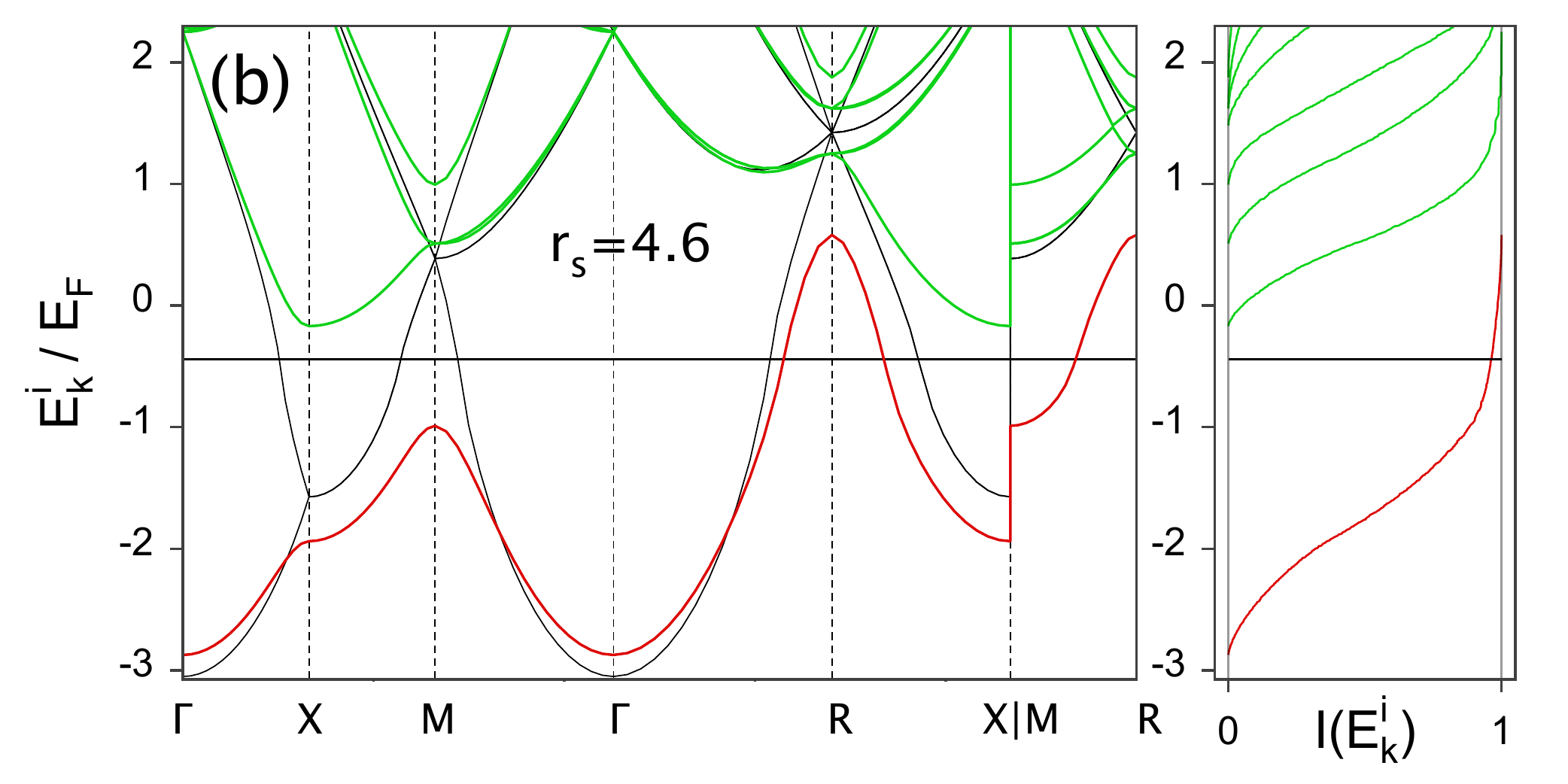}
\includegraphics[scale=0.42]{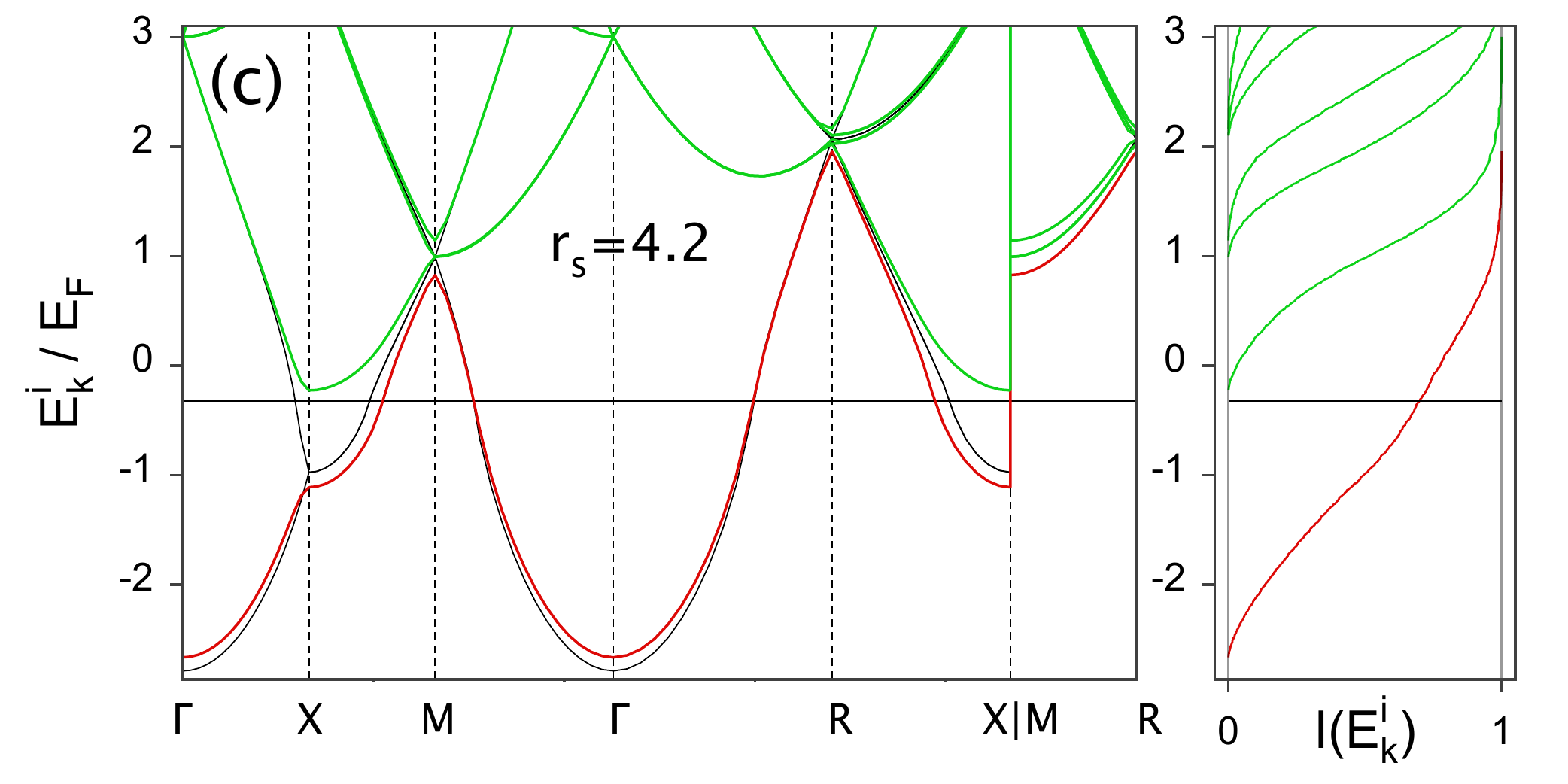}
\includegraphics[scale=0.42]{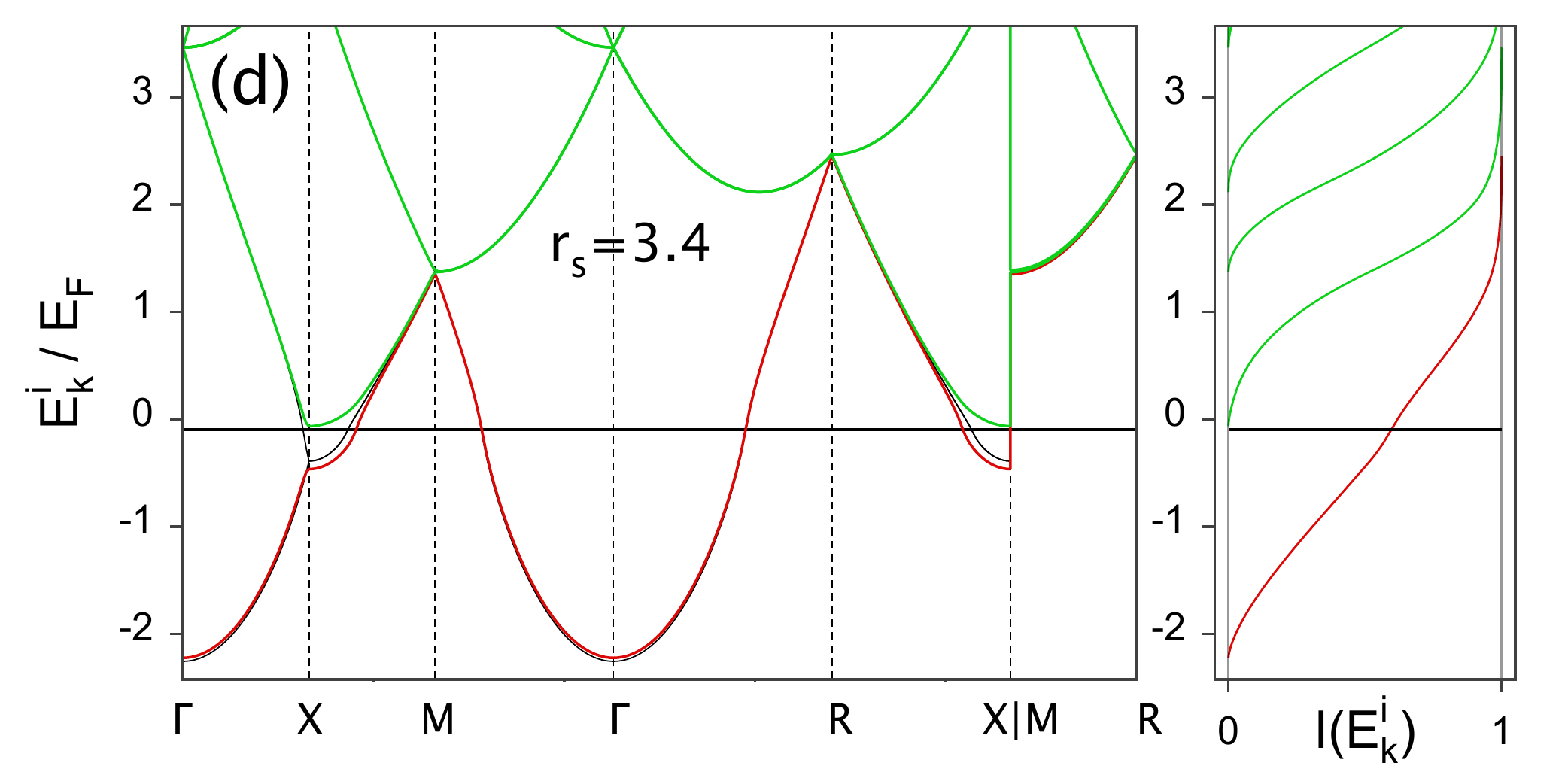}
\caption{HF band structure along standard path\cite{BZ-Path} (see Table \ref{TAB-LATTICE}) of U sc at various density for $Q=Q_W=1.612$ (a),1.631 (b), 1.817 (c) and 1.918 (d). 
Red, green and black lines stand for first band, higher bands, and FG energies, respectively.
In each picture, the right plot shows the energy versus the integrated density of state of each band.
The full horizontal line represents the last occupied states.
In (a), the gap is the domain between this full line and the dashed line. In (b), (c) and (d), the gap vanishes.
\label{FIG-GAP}
}
\end{center}
\end{figure*}

From the eigenvalues of the HF Hamiltonian, Eq.\ref{eq-defHF}, we can plot the band structure at different densities.
The band structure of the unpolarized gas in the sc symmetry for some $r_s$ is shown in Fig.\ref{FIG-GAP}. 
For {\it high} $r_s$ (Fig.\ref{FIG-GAP}-a), the solution is a WC: the first band is occupied with a finite gap corresponding to an insulator.
At lower $r_s$ (Fig.\ref{FIG-GAP}-b,c,d), the solution becomes IC and the first band is partially occupied, corresponding to a metal.
As $r_s$ decreases, the bands  approach the FG band structure.
For other symmetries or polarizations, the scenarios are very similar.

\begin{figure}
\begin{center}
\includegraphics[scale=0.42]{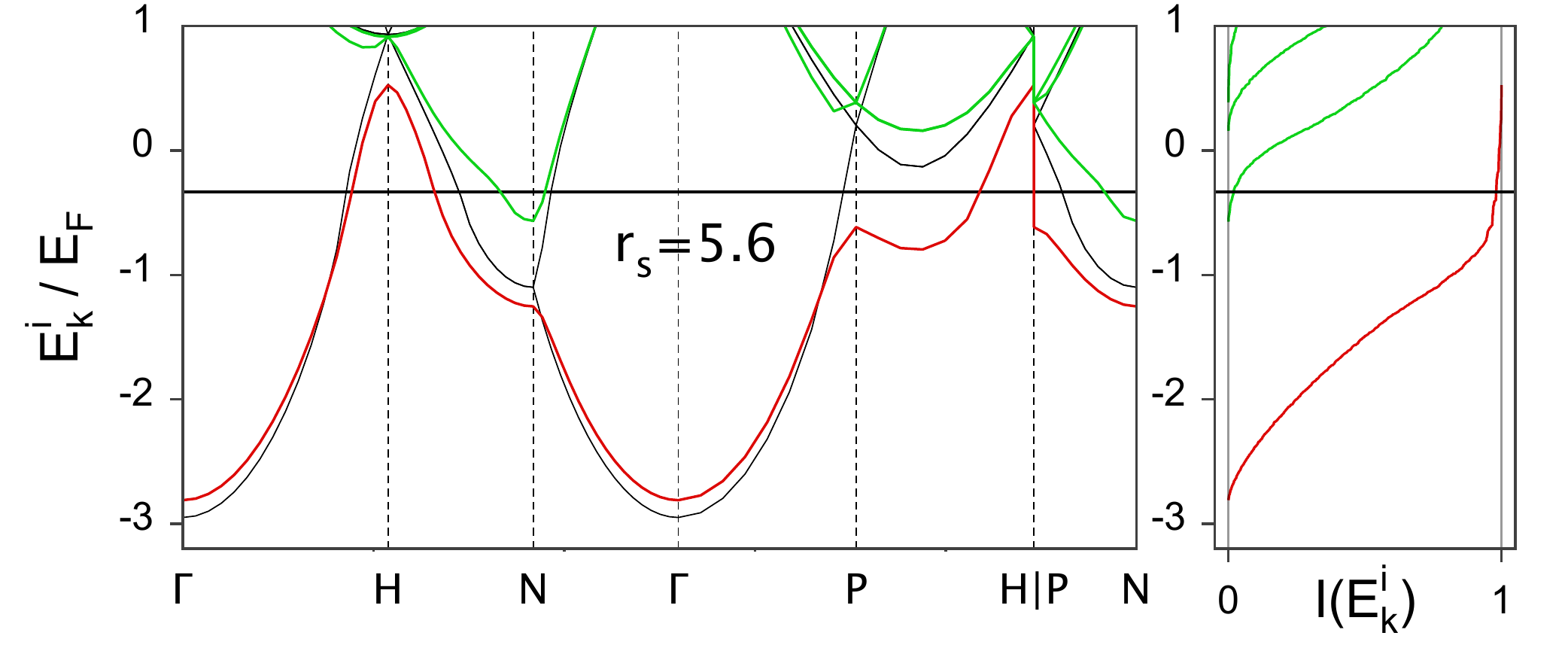}
\caption{Same as Fig.\ref{FIG-GAP} for P bcc at $Q=Q_W$ with several bands. This solution has a lower energy than WC (one band) and FG. However, at the same density a better solution is found with $Q>Q_W$.
\label{FIG-GAP-QW}
}
\end{center}
\end{figure}

Imposing $Q=Q_W$ at low $r_s$ leads to solution with several partially occupied bands, as shown in Fig.\ref{FIG-GAP-QW}, and the Fermi gas can be recovered (still at $Q=Q_W$). However, when this happens, solutions with lower energies are found with $Q>Q_W$ leading to IC states.

\subsection{Ground state wave function characterization}
\label{SEC-Characterization}

\begin{figure}[tb]
\begin{center}
\includegraphics[scale=0.5]{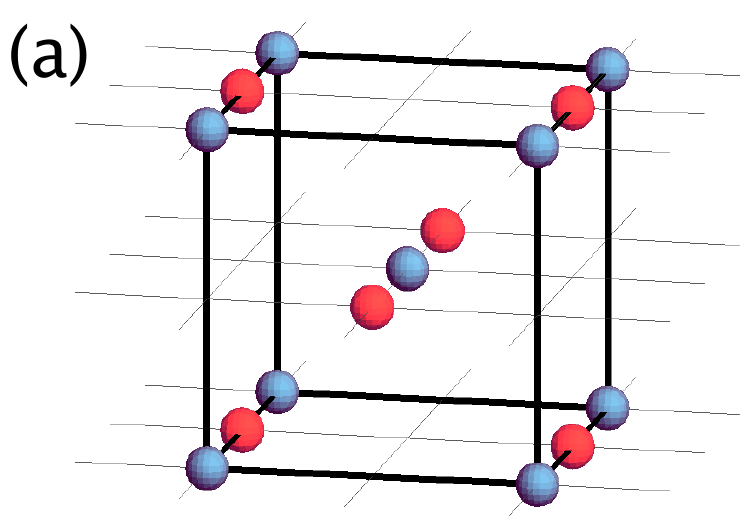}
$\quad$ 
\includegraphics[scale=0.5]{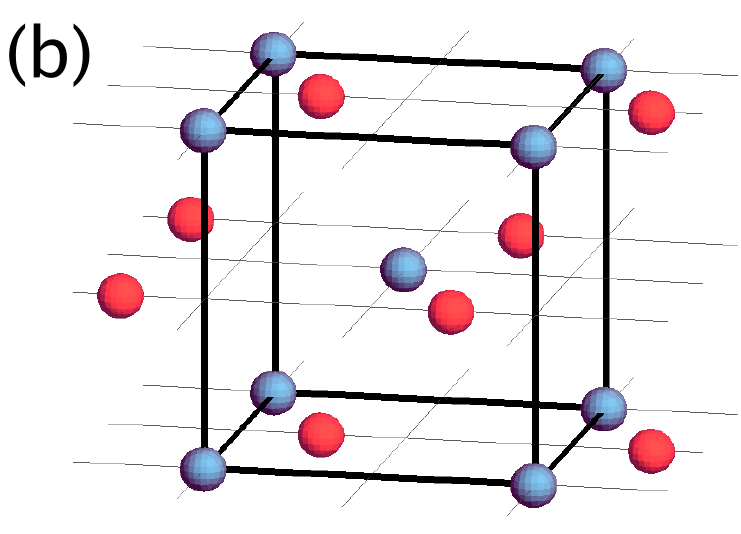}
\caption{U bcc real space maxima. 
Blue and red points stand for up and down spin density, respectively. 
(a) : $r_s\gtrsim5.6$ WC states where up and down spins are {\it aligned}.
(b) : $r_s\lesssim5.6$ IC and WC states where up and down spins are {\it tilted} 
\cite{WEB}.
\label{FIG-RHOR-BCC}
}
\end{center}
\end{figure}

At fixed $r_s$, geometry, and polarization, only one band is occupied (see Sec.\ref{SEC-GS-FIXEDSYM}). 
Thus, the occupied Bloch waves may be indexed by $\bk$ and $\sigma$
\begin{align}
	\ket{u_{\bk,\sigma}}&=\sum_{\bq\in\Lambda}a_{\bk,\sigma}(\bq)\ket{\bk+\bq \,; \sigma}
\end{align}
where $\bk$ belongs to a subset of $\B$ (or exactly $\B$ for the WC).
For P states, we find that $a_{\bk\uparrow}(\bq)$ can be chosen real positive.
For  most U states, we find that spin up follow the same rule while spin down have an additional phase described by a simple translation $\bT=(\alpha_0,\alpha_1,\alpha_2)$, in real space lattice basis, where $\alpha_i=0$ or $1/2$: $a_{\bk\downarrow}(\bq)=a_{\bk\uparrow}(\bq)e^{i\bq.\bT}=\pm a_{\bk\uparrow}$.
The ground states of U sc and fcc for all $r_s$ are described by such a wave function,
as well as incommensurate Hex and the WC U bcc at $r_s>5.6$ (see Fig.\ref{FIG-RHOR-BCC}-a).

In the second family of ground states, we still have $\|a_{\bk\uparrow}(\bq)\|=\|a_{\bk\downarrow}(\bq)\|$.
This family contains the WC with Hex symmetry and the incommensurate and WC phase with bcc symmetry at $r_s<5.6$ (see Fig.\ref{FIG-RHOR-BCC}-b.\cite{WEB})

In the \HT case, there are two electrons per unit cell so the system occupies two bands and the amplitudes are complex numbers.

\subsection{Correlation functions}
\label{SEC-Gofr-Sk}
We define the 1-body density $\rho_1$ as:
\begin{align}
\nonumber
	\rho_1(\br,\sigma&;\br',\sigma')=\sum_{\bk,\bk'} \rho_1(\bk,\sigma;\bk',\sigma')e^{i(\bk\br-\bk'\br')}\\
	&=\sum_{\substack{\bk\in\B\\\bq',\bq'\in\Lambda}}\rho_\bk(\bq,\sigma;\bq',\sigma')e^{i\bk(\br-\br')}e^{i(\bq\br-\bq'\br')}
\end{align} 
We consider the two-body correlation function, using the same notation of Eq.(\ref{eq-rho2-rho1}) 
\begin{align}
	\chi(\uo,\ut;\uo',\ut')=\rho_2(\uo,\ut;\uo',\ut')-\rho_1(\uo;\uo')\rho_1(\ut;\ut')
\end{align} 
From the HF factorization of the two-body density matrix, Eq.~\ref{eq-rho2-rho1}, we obtain the correlation function $\chi$, defined such that $\chi=0$ when the correlations vanish:
\begin{align}
	\chi_{\sigma,\sigma'}(\br,\br')=-\rho_1(\br,\sigma;\br',\sigma')\rho_1(\br',\sigma';\br,\sigma)
\end{align} 
and, concerning the average distance of a pair,
\begin{align}
\nonumber
	\chi_{\sigma,\sigma'}&(\br)=\lim_{V\rightarrow \infty}\frac 1V\int  d\br'\,\chi_{\sigma,\sigma'}(\br'+\br,\br')\\
\label{EQ-chir}
	&=-\sum_{\bq\in\Lambda}\left| \sum_{\substack{\bk\in\B\\\bq'\in\Lambda}} \rho_\bk(\bq',\sigma;\bq'-\bq ,\sigma')e^{i(\bk+\bq)\br}\right|^2
\end{align} 
and its Fourier transform is denoted $\tilde \chi_{\sigma,\sigma'}(\bk)$.
For unpolarized systems, we have $\chi_{\upuparrows}=\chi_{\downdownarrows}$ and $\chi_{\uparrow\downarrow}=0$.

\begin{figure}
\begin{center}
\includegraphics[scale=0.5]{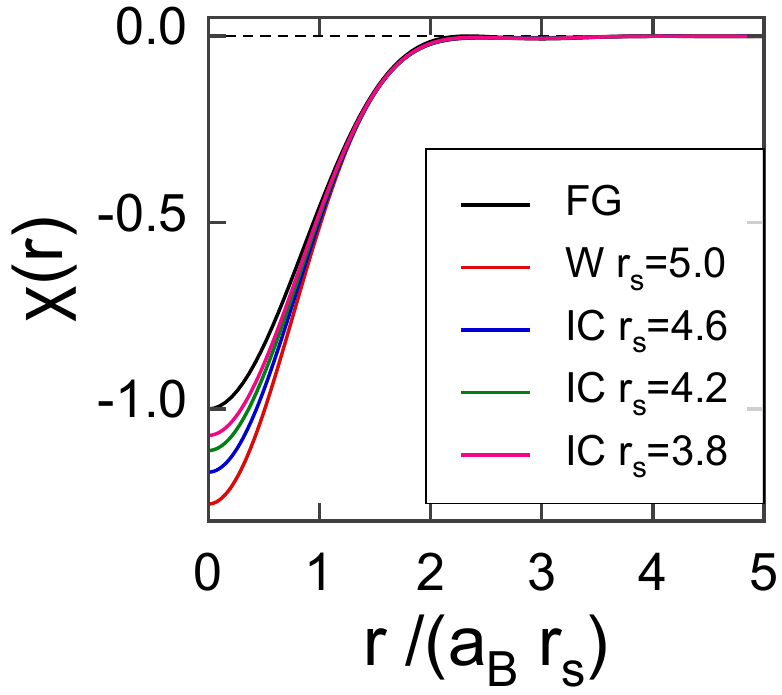}
\includegraphics[scale=0.5]{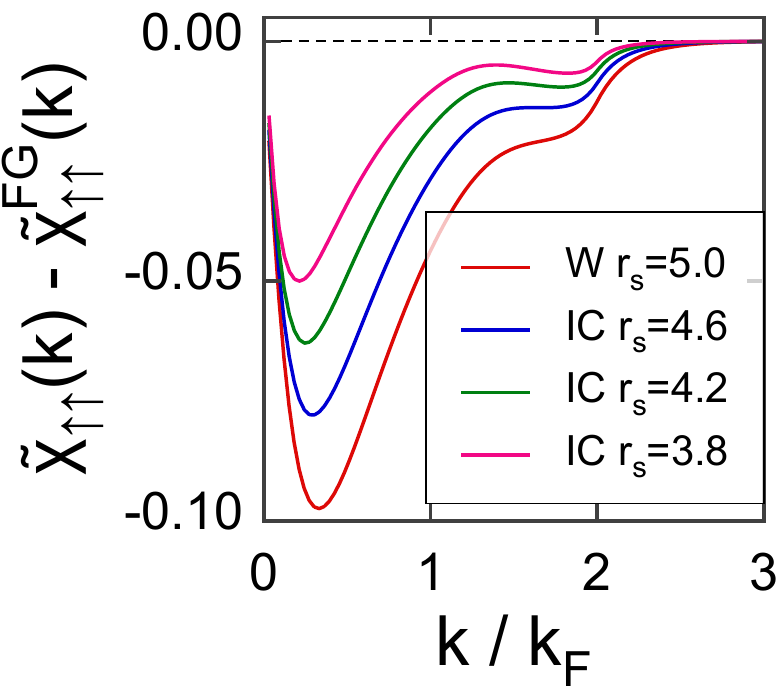}
\caption{Angle-averaged pair correlation function for the unpolarized gas in bcc symmetry, 
$\chi(r)$, see Eq. (\ref{EQ-chir}), and the difference between its Fourier transform $\tilde \chi(k)$ and ${\tilde \chi}^{\rm FG}(k)$ for $r_s=5$, 4.8, 4.2 and 3.8 at $Q=1.809$, 1.818, 1.827 and 1.836 respectively.
At smaller values of $r_s$, inside the IC phase with $Q>1.9$, the difference to the FG is  less than $10^{-2}$.
\label{FIG-GR-SK}
}
\end{center}
\end{figure}

Figure \ref{FIG-GR-SK} shows the comparison of the HF-correlation functions with the Fermi gas solution,
$\chi^{FG}_{\upuparrows}(r)=-9\left[\sin(x)-x\cos (x)\right]^2/{x^6}$ with $x=k_Fr$ and 
$\tilde\chi^{FG}_{\upuparrows}(k)=3q/4 - q^3/16$ with $q=k/k_F<2$ 
and $\tilde\chi^{FG}_{\upuparrows}(k)=0$ for $k>2k_F$. 
As one can see, the pair correlations change smoothly between different phases and remain close the FG.

\section{Conclusion}
\label{SEC-Conclusion}

In this paper we have presented our numerical algorithm which has been used to obtain the HF phase diagram 
presented in Ref.\cite{HF3DEG-letter}, and discussed various properties of the solutions, focusing on the IC phase.
 Here, we have given evidence that the IC
states are very close to pure spin or charge density wave superimposed in different directions, so that the 
structure proposed by Overhauser\cite{Overhauser} in 1962 is essentially recovered.
From the band structure, we expect these states to have metallic character very close to the usual FG and also similar
pair correlations.\cite{WEB}

Is the spin density ground state stable against the inclusion of correlations beyond the Hartree-Fock approximation?
In the high density region, correlation energies \cite{CA} largely exceeds the HF energy difference between 
spin or charge density waves and the homogeneous FG. Nevertheless, since these broken symmetry states only introduce tiny modifications in the very proximity of the spherical Fermi surface of the FG, correlations will shift their energies by almost the same amount
as the FG. Assuming a rigid shift in the correlation energy, our calculations indicate the possibility of spin or charge density waves at rather low temperatures. In real materials, they will then be in competition to other instabilities, e.g. superconductivity.

\begin{table*}
\begin{center}
\renewcommand{\arraystretch}{1.5}
\begin{tabular}{c||c|c|c|cc}
symmetry
& sc
& bcc
&fcc
& Hex
& Hex$^{(2)}$
\\
\hline

$n_c$&1&1&1&1&2\\

$M_Q$
& $\left(\begin{array}{rrr}
	1 & 0 & 0 \\
	0 & 1 & 0 \\
	0 & 0 & 1
	\end{array}\right)$
& $\scalebox{1.4}{$\frac{1}{\sqrt{2}}$}\left(\begin{array}{rrr}
	0 & 1 & 1 \\
	1 & 0 & 1 \\
	1 &1 & 0
	\end{array}\right)$
& $\scalebox{1.4}{$\frac{1}{\sqrt{3}}$}\left(\begin{array}{rrr}
	-1 & 1 & 1 \\
	1 & -1 & 1 \\
	1 & 1 & -1
	\end{array}\right)$
& \multicolumn{2}{c}{$\scalebox{1.4}{$\frac{4\sqrt{2}}{3}$}\left(\begin{array}{ccc}
	1 & -0.5 & 0 \\
	0 & \sqrt{3}/2 & 0 \\
	0 & 0 & 3/(4\sqrt{2})
	\end{array}\right)$}
\\
$\gamma^3$&1&$\frac1{\sqrt2}$&$\frac4{3\sqrt3}$&\multicolumn{2}{c}{$\frac{16}{3\sqrt3}$}\\
$Q_W/k_F$ 
& $1.611991954016$
& $1.809399790564$
& $1.758882522024$
& $1.108026556895$
& $0.879441261012$
\\
$d^2$
& $n_1^2+n_2^2+n_3^2$
& $d^2_{sc}+n_1n_2+n_1n_3+n_2n_3$
& $d^2_{sc}-\frac{2}{3}(n_1n_2+n_1n_3+n_2n_3)$
& \multicolumn{2}{c}{$\frac{32}{9} (n_1^2+n_2^2-n_1n_2)+n_3^2$}
\\
$M_\Lambda$
& 7 , 19 , {\bf 27} , 33 , 57
& 13 , {\bf 19} , 43 , 55 , 79
& 9 , {\bf 15} , 27 , 51 , 59
& \multicolumn{2}{c}{3 , 9 , 11 , {\bf 23} , 35}
\\
$C_\Lambda$
& -2.837297479481
& -2.888461503054
& -2.888282119020
& \multicolumn{2}{c}{-2.512880623796}
\\ \hline
S-P
& \parbox{2cm}{
\begin{align*}\begin{array}{c@{\ : \ (}ccc@{)}}
 \Gamma &  0 & 0 & 0 \\
 X &  \frac12 & 0 & 0 \\
 M & \frac12 & \frac12 & 0 \\
 R & \frac12 & \frac12 & \frac12
 \end{array}\end{align*}
 }
& \parbox{2cm}{
\begin{align*}\begin{array}{c@{\ : \ (}l@{\ \ \ }cr@{)}}
 \Gamma &  0 & 0 & 0 \\
 N &  \frac12 & 0 & 0 \\
 H & \frac12 & \frac12 & -\frac12 \\
P & \frac14 & \frac14 & \frac14
 \end{array}\end{align*}
 }
& \parbox{2cm}{
\begin{align*}\begin{array}{c@{\ : \ (}c@{\ \ \ }cc@{)}}
 \Gamma &  0 & 0 & 0 \\
 X &  \frac12 & \frac12 & 0 \\
 W & \frac12 & \frac34 & \frac14 \\
 K & \frac38 & \frac34 & \frac38 \\
 L & \frac12 & \frac12 & \frac12 \\
U & \frac58 & \frac58 & \frac14
 \end{array}\end{align*}
 }
&  \multicolumn{2}{c}{\parbox{2cm}{
\begin{align*}\begin{array}{c@{\ : \ (}c@{\ \ \ }cc@{)}}
 \Gamma &  0 & 0 & 0 \\
 M &  \frac12 & 0 & 0 \\
 K & \frac23 & \frac13 & 0 \\
 A & 0 & 0 & \frac12 \\
  L & \frac12 & 0 & \frac12 \\
 H & \frac23 & \frac13 & \frac12
 \end{array}\end{align*}
 }}
\end{tabular}
\caption{Lattice definitions and properties. $n_c$ is the number of sites per primitive cell. Matrix $M_Q$ is defined by $M_Q=(\bQ_1/Q, \bQ_2/Q, \bQ_3/Q)$  where $\bQ_i$ are normalized reciprocal lattice vectors in cartesian coordinates. For the hexagonal case, $Q=||\bQ_3||$. $\gamma^3Q^3$ is the volume of $\B$. $Q_W/k_F=\gamma(4\pi/(3n_c))^{1/3}$. $d^2$ is the square distance in the basis $\bQ_i$. 
$M_\Lambda$'s values are the numbers of vectors with integer coordinates in the first shells. 
Bold values indicate the minimum of $M_\Lambda$ for which a neighborhood of $\B$ is covered.
$C_\Lambda$ is the Madelung constant. S-P are symmetry points of $\B$\cite{BZ-Path}, given in reciprocal lattice vectors coordinates.}
\label{TAB-LATTICE}
\end{center}
\end{table*}

\end{document}